\documentclass[a4paper,12pt]{article}
\usepackage{mathrsfs}
\usepackage{graphicx} 
\usepackage{epstopdf}
\usepackage{subfigure}

\begin{document}

\hoffset = -1truecm \voffset = -2truecm \baselineskip = 10 mm

\title{\bf The current recorded signals of ultrahigh-energy $\gamma$-rays may come from EeVatrons in the galaxy}

\author{Wei Zhu$^a$, Peng Liu$^b$, Zhiyi Cui$^a$ and Jianhong Ruan$^a$
        \\
        \normalsize $^a$Department of Physics, East China Normal University,
        Shanghai 200241, China \\
        \normalsize $^b$School of software, Shanxi Agricultural University, Shanxi Jinzhong 030801, China\\
    }
\date{}

\newpage

\maketitle

\vskip 3truecm

\begin{abstract}

    A hard $\gamma$-ray spectrum of supernova remnant G106.3+2.7
can be explained by using the hadronic model with the gluon condensation effect.
This implies that not only PeVatrons but also EeVatrons
generally exist in the universe including
our galaxy, and they can accelerate protons to beyond "ankle" ($10^{19}~eV$).
Although these proton beams are very weak in the galaxy and cannot be
observed individually on the earth, the gluon condensation effect may
greatly enhance the proton-proton cross section, which can compensate for the weak proton flux and produce the
observed $\gamma$-rays. We also show that the gluon condensation effect in proton provides an
efficient conversion mechanism for kinetic energy into $\gamma$-rays in the universe.

\end{abstract}

{\bf keywords}: Cosmic ray spectra; EeVatron; Gluon condensation

\newpage

\vskip 1truecm
\section{Introduction}

   The energy distributions of cosmic rays (mainly protons) are well
described by a power law $E_p^{-\beta_p}$.  It is generally believed that the cosmic rays under
"knee" ($E_p< 10^{15}~eV$) can be
generated by astrophysical sources in our galaxy, while above the "knee"
by the extragalactic origins.  Recently $\gamma$-rays from the galaxy beyond 100 $TeV$, even up to 1.4 $PeV$ have been discovered [1,2]. For example, the Tibet AS$\gamma$ collaboration reports
the observation of $\gamma$-ray around 100 $TeV$ from
the supernova remnant G106.3+2.7 .
According to the standard hadronic mechanism, these events imply that $PeV$ accelerator
(PeVatron) should generally exist in the galaxy.

    The $\gamma$-rays are produced through $p+p\rightarrow \pi^0\rightarrow 2\gamma$ in the hadronic scenario
in which the spectra of $\gamma$-rays at the rest frame of pion have a maximum value at $E_{\gamma}=m_{\pi}/2$,
and the parent proton flux $\Phi_p\sim E_p^{-\beta_p}$ with $\beta_p\simeq 2.7$.
Using the parameterized relation between $N_{\pi}$ (pion number) and $E_p$, this mechanism predicts
that the $\gamma$-ray distribution $E^2_{\gamma}\Phi_{\gamma}$ has a so-called ``$\pi^0$-decay bump'' at $E_{\gamma}\sim 1~GeV$,
which has been confirmed by SNRs IC443 and W44 [3].

    However, the peak of $E^2_{\gamma}\Phi_{\gamma}$ in
SNR G106.3+2.7 is localized near $20~TeV$ rather than 1 $GeV$. For fitting the data,
a special proton spectrum $\Phi_p\sim E_p^{-\beta_p}\exp(-E_p/E_p^{cut})$ is needed,
where hard index $\beta_p=1.8$ and the cut energy $E_p^{cut}=0.5~PeV$. The latter parameter is an evidence of existing PeVatron in the galaxy. However, what puzzling is that, we still have not yet found the hardening cosmic ray spectra since $\beta_p=1.8<2.7$.
If we assume that the flux of these accelerated protons in PeVatron is so weak, then how can we explain the
recorded radiation signals?
One possible answer is
that the cross section of $p+p\rightarrow \pi$ is abnormally increased in the observed high energy band,
which can compensate for the weak proton flux, and generate the measurable $\gamma$-rays.

    In this work we report that the gluon condensation effect in proton is just such a mechanism.
A series of QCD researches predict that huge numbers of gluons in the proton may gather at a critical momentum
[4-6]. We call it the gluon condensation (GC).
We know that gluons dominate the relativistic proton-proton inelastic collisions, where they convert the kinetic energy of parent
protons into pions and subsequent photons.
Once these condensed gluons are excited and participate
to the $pp$ collision, they will greatly enhance the $pp$ cross section, and generate a lot of $\gamma$-rays
to compensate the weak parent proton flux. We emphasize that these
$\gamma$-ray spectra have a GC-character and can be recognized in the experimental data.
After comparisons of the GC model with the AS$\gamma$ data we find that
not only PeVatrons but also EeVatrons
generally exist in the universe including
our galaxy, and they can accelerate the protons to beyond ``ankle'' ($>10^{19}~eV$).
Besides, the GC effect in proton also presents the most
efficient conversion of kinetic energy of parent protons to $\gamma$-ray radiation.

\section{Hadronic scenario with the GC effect}

     According to the hadronic model of radiation, about half energies of parent protons are taken away
by the valence quarks, which form the leading particles, and the remaining energies are transformed into the
secondary hadrons (mainly pions) in central region through gluons.
The flux of high energy $\gamma$-rays in laboratory frame reads [7-11]

$$\Phi_{\gamma}(E_{\gamma})=C_{\gamma}\left(\frac{E_{\gamma}}{1GeV}\right)^{-\beta_{\gamma}}
\int_{E_{\pi}^{min}}^{E_{\pi}^{max}}dE_{\pi}
\left(\frac{E_p}{1GeV}\right)^{-\beta_p}$$
$$\times N_{\pi}(E_p,E_{\pi})
\frac{d\omega_{\pi-\gamma}(E_{\pi},E_{\gamma})}{dE_{\gamma}},
\eqno(2.1)$$ where the spectral index $\beta_{\gamma}$
denotes the propagation loss of $\gamma$-rays. The accelerated protons obey a power law
$N_p\sim E_p^{-\beta_p}$ in the source.
$C_{\gamma}$ incorporates the kinematic factor and the flux
dimension.
Neglecting the harmonization mechanism, $N_{\pi}$ is proportional to
the cross section of gluon mini-jet production [6]

$$\frac{d\sigma_g}{dk_T^2dy}=\frac{64N_c}{(N^2_c-1)k_T^2}\int q_T d
q_T\int_0^{2\pi}
d\phi\alpha_s(\Omega)\frac{F(x_1,\frac{1}{4}(k_T+q_T)^2)F(x_2,\frac{1}{4}
(k_T-q_T)^2)}{(k_T+q_T)^2(k_T-q_T)^2},
\eqno(2.2)$$where
$\Omega=Max\{k_T^2,(k_T+q_T)^2/4,
(k_T-q_T)^2/4\}$; the longitudinal momentum
fractions of interacting gluons are fixed by kinematics
$x_{1,2}=k_Te^{\pm y}/\sqrt{s}$.

    A QCD study predicts
that gluons in proton may converge at a critical momentum [4-6].
The GC should induce significant effects in
proton collision if the proton energy exceeds the GC-threshold. The energy of proton accelerated inside the sources, such as
supernova remnants (SNRs), active galactic nuclei (AGN) or pulsars,
could reach a very high level. In general, the more gluons, the more
secondary pions. One can
image that the pion yield in this case reaches its maximum
value due to the GC effect, i.e., almost all available kinetic energies of collisions at the
center-of-mass system are used to create pions.
Taking this approximation, one can avoid the complicated hadronization mechanism
and use the relativistic invariance and energy conservation to
directly obtain the solution $N_{\pi}$ in the $pp$ collision

$$\ln N_{\pi}=0.5\ln E_p+a, ~~\ln N_{\pi}=\ln E_{\pi}+b,  \eqno(2.3)$$
$$~~ where~E_{\pi}
\in [E_{\pi}^{GC},E_{\pi}^{max}], $$where $a\equiv 0.5\ln (2m_p)-\ln m_{\pi}+\ln K$
and $b\equiv \ln (2m_p)-2\ln m_{\pi}+\ln K$. $K\simeq1/2$ is inelasticity.
Equation (2.3) gives the one-by-one relations between $N_{\pi}$, $E_p$ and
$E_{\pi}^{GC}$. Particularly, within the range of the GC-effect we have

$$E_p=\frac{2m_p}{m^2_{\pi}}E_{\pi}^2.\eqno (2.4)$$Substituting (2.3) and the standard spectrum of $\pi^0\rightarrow 2\gamma$ into (2.1)
one can get the GC-characteristic spectrum

$$E_{\gamma}^2\Phi^{GC}_{\gamma}(E_{\gamma})\simeq\left\{
\begin{array}{ll}
\frac{2e^bC_{\gamma}}{2\beta_p-1}(E_{\pi}^{GC})^3\left(\frac{E_{\gamma}}{E_{\pi}^{GC}}\right)^{-\beta_{\gamma}+2} \\ {\rm ~~~~~~~~~~~~~~~~~~~~~~~~if~}E_{\gamma}\leq E_{\pi}^{GC},\\\\
\frac{2e^bC_{\gamma}}{2\beta_p-1}(E_{\pi}^{GC})^3\left(\frac{E_{\gamma}}{E_{\pi}^{GC}}\right)^{-\beta_{\gamma}-2\beta_p+3}
\\ {\rm~~~~~~~~~~~~~~~~~~~~~~~~ if~} E_{\pi}^{GC}<E_{\gamma}<E_{\pi}^{cut},\\\\
\frac{2e^bC_{\gamma}}{2\beta_p-1}(E_{\pi}^{GC})^3\left(\frac{E_{\gamma}}{E_{\pi}^{GC}}\right)^{-\beta_{\gamma}-2\beta_p+3}
\exp\left(-\frac{E_{\gamma}}{E_{\pi}^{cut}}+1\right).
\\ {\rm~~~~~~~~~~~~~~~~~~~~~~~~ if~} E_{\gamma}\geq E_{\pi}^{cut},
\end{array} \right. \eqno(2.5)$$or

$$\Phi^{GC}_{\gamma}(E_{\gamma})\equiv\left\{
\begin{array}{ll}
\Phi_0\left(\frac{E_{\gamma}}{E_{\pi}^{GC}}\right)^{-\Gamma_1} \\ {\rm ~~~~~~~~~~~~~~~~~~~~~~~~if~}E_{\gamma}\leq E_{\pi}^{GC},\\\\
\Phi_0\left(\frac{E_{\gamma}}{E_{\pi}^{GC}}\right)^{-\Gamma_2}
\\ {\rm~~~~~~~~~~~~~~~~~~~~~~~~ if~} E_{\pi}^{GC}<E_{\gamma}<E_{\pi}^{cut},\\\\
\Phi_0\left(\frac{E_{\gamma}}{E_{\pi}^{GC}}\right)^{-\Gamma_2}
\exp\left(-\frac{E_{\gamma}}{E_{\pi}^{cut}}+1\right),\\
\\ {\rm~~~~~~~~~~~~~~~~~~~~~~~~ if~} E_{\gamma}\geq E_{\pi}^{cut}.
\end{array} \right. \eqno(2.6)$$Note that the GC-threshold $E_{\pi}^{GC}$
is target $A$-dependent in the $p-A$ (or $A-A$) collisions, while $E_{\pi}^{cut}$ relates to the accelerator properties.
The double break power law (2.6) is a feature of the GC effect in high energy $\gamma$-ray spectra.

\section{Explanation of the gamma spectra of G106.3+2.7}

    We fit the spectral energy distribution of SNR G106.3+2.7 [1] combining the data of Fermi-LAT and
VERITAS [12,13] in figure 1.
The fitting quality for $\gamma$-ray spectrum from 5 $GeV$ to 100 $TeV$ is $\chi^2/d.o.f.=15/(17-4)=1.15$.
The GC-spectrum (2.5) predicts a single power law in $E_{\pi}^{GC}<E_{\gamma}<E_{\pi}^{cut}$.
The $AS\gamma$ collaboration reports that its $\gamma$-ray energy spectrum from 6 $TeV$ to 115 $TeV$ can also be fitted by the single power law
with $\chi^2/d.o.f=2.5/(7-2)=0.5$,
which is perfectly consistent with our prediction.

     The AS$\gamma$ data show that $E_{\pi}^{cut}>100~TeV$. According to (2.4) we have $E_p>10^3$ $EeV$.
This conclusion is obviously different from the standard hadronic model without the GC effect. In the latter case,
$E_p^{max}\sim 10\times E_{\gamma}^{cut}$, thus $E_p^{max}>1$ $PeV$ since $E_{\gamma}^{cut} > 100$ $TeV$ for G106.3+2.7. While the GC model predicts
a different relation (2.4).
This is not surprising, since the extremely high energies of the parent protons are needed to generate a lot of $\gamma$-rays near $E_{\pi}^{GC}\sim 20$ $TeV$ through the GC-effect due to the total energy conservation.
In this process the condensed gluons in proton play the key role in the
conversion from collision energy to secondary particles. Interestingly, we have found that the spectrum of Tycho's supernova remnant has $E_{\pi}^{GC}=400$ $GeV$ and $E_{\pi}^{cut}>5$ $TeV$ [9],
which means
that protons can be accelerated above $E_p=10^{18}$ $eV$ in SRN Tycho. Therefore, we consider that both PeVatron and EeVatron generally exist in the universe including our galaxy.

    We have not yet directly measured these ultra-high-energy protons in our galaxy since their fluxes are too weak. The recorded
primary cosmic ray spectra beyond $10^{15}~eV$ are actually
superpositions of countless proton fluxes from the extragalactic sources. Its spectrum has been softened due to
diffusion and absorption through long-distance propagation in interstellar medium.

    A big difference in the explanations of ultrahigh-energy $\gamma$-ray spectra between the standard hadronic model and the GC model
originates from the different understandings of the high energy $pp$ collision. The former takes the experience cross section at $TeV$ band
in the Large Hadron Collider (LHC), while the later considers the GC-contributions at the $PeV$ band.

    The break position $E_{\pi}^{GC}$ is determined by the gluon distribution in different nuclei, for example,
$E_{\pi}^{GC}(p-A)>E_{\pi}^{GC}(p-A')$ if $A<A'$ since the nonlinear corrections enhance as $A$ increase. Therefore,
there is a maximum value of $E_{\pi}^{GC}$ corresponding to the $pp$ collision where $A=1$. We assume that $E_{\pi}^{GC}(p-p)
\simeq 20~TeV$ since it is a maximum value of $E_{\pi}^{GC}$ recorded so far (figure 1).

    This work also shows that the GC effect is one of the most powerful converter
for proton kinetic energy to $\gamma$-rays at high energy in the universe. Although electron-positron annihilation
and inverse Compton scattering can also be such converter, their actual efficiencies are restricted by low densities of positrons and soft photons.

\section{Conclusions}

 Charged particles (mainly protons) in cosmic rays are accelerated to ultra-high-energy in the universe by
known and unknown mechanisms. Generally believed that our galaxy can only generate protons with energy
lower than $10^{15}~eV$, and higher-energy protons come from
extragalactic sources. However, according to the hadronic model with the GC-effect, the $\gamma$-ray spectrum of SNR G106.3+2.7
challenges the above points of view, and
the new discoveries of AS$\gamma$ and LHAASO lead us to believe that
EeVatrons exist in \sl all\rm~ galaxies, which can accelerate protons to
$10^{19}~eV$ and beyond. In addition, the GC effect in proton shows the most powerful converter
for proton kinetic energy to $\gamma$-rays.
Although the acceleration mechanism of cosmic rays is not the content of this work,
the above conclusions may provide a new perspective to find EeVatron in the galaxy.

\noindent {\bf Acknowledgments}:

    We thank K. Murugan for useful comments. This work is supported by the National Natural
Science of China (No.11851303).


\newpage
\begin{thebibliography}{99}

\bibitem{1} M. Amenomori, Y.M. Bao and X.J. Bi et al. (The Tibet AS$\gamma$ collaboration)
\begin{sl}Potential PeVatron supernova remnant G106.3+2.7 seen in the highest-energy gamma rays.
Nature Astron.\end{sl} (2021) doi:10.1038/s41550-020-01294-9.

\bibitem{2} C. Zhen, F.A. Aharonia and Q. An et al. (The LHAASO collaboration) \begin{sl}Ultrahigh-energy
photons up to 1.4 petaelectronvolts
from 12 $\gamma$-ray Galactic sources.  Nature.\end{sl} (2021) doi.org/10.1038/s41586-021-03498-z.

\bibitem{3} M. Ackermann, M, Ajello and A. Allafort et al. \begin{sl}Detection of the characteristic
pion-decay signature in supernova remnants. Science\end{sl} $\bf{339}$ (2013) 8074.

\bibitem{4} W. Zhu, Z.Q. Shen and J.H. Ruan, \begin{sl}Can a chaotic solution in the QCD evolution equation restrain
high-energy collider physics? Chin Phys Lett\end{sl} $\bf{25}$ (2008) 3605.

\bibitem{5} W. Zhu, Z.Q. Shen and J.H. Ruan, \begin{sl}The chaotic effects in a nonlinear QCD evolution equation.
Nucl. Phys. \end{sl} $\bf{B911}$ (2016) 1.

\bibitem{6} W. Zhu and J.S. Lan, \begin{sl}The gluon condensation at high energy hadron collisions
Nucl. Phys.\end{sl} $\bf{B916}$ (2017) 647.

\bibitem{7} W. Zhu, J.S. Lan and J.H. Ruan, \begin{sl}The gluon condensation in high energy cosmic rays
\sl Int J Mod Physics.\end{sl} $\bf{E27}$ (2018) 1850073.

\bibitem{8} F. Feng, J.H. Ruan, F. Wang and W. Zhu, \begin{sl}Looking for the Gluon condensation signature in protons using the Earth-limb
gamma-ray spectra. Astrophys. J.\end{sl} $\bf{868}$ (2018) 2.

\bibitem{9} W. Zhu, P. Liu, J.H. Ruan and F. Wang, \begin{sl}Possible evidence for the gluon condensation
effect in cosmic positron and gamma-ray spectra, Astrophys. J.\end{sl} $\bf{889}$ (2020) 127.

\bibitem{10} W. Zhu, Z.C. Zheng, P. Liu, L.H. Wan, J.H. Ruan and F. Wang, \begin{sl}Looking for the possible gluon
condensation signature in sub-TeV gamma-ray spectra: from active galactic nuclei to gamma ray bursts,
JCAP.\end{sl} $\bf{01}$ (2021) 038.

\bibitem{11} W. Zhu, P. Liu, J.H. Ruan, R.Q. Wang and F. Wang, \begin{sl}The gluon condensation effect in the
cosmic hadron spectra, JCAP.\end{sl} $\bf{09}$ (2020) 011.

\bibitem{12} Y. Xin, H. Zeng, S. Liu, Y. Fan and D. Ver Wei, \begin{sl}J2227+608: a hadronic PeVatron pulsar wind nebula?
Astrophys J, \end{sl} $\bf{885}$ (2019) 162.

\bibitem{13} V.A. Acciari et al. \begin{sl}Detection of extended VHE gamma ray emission from G106.3+2.7 with
VERITAS, Astropys. J.\end{sl} $\bf{703}$ (2009) L6.


\newpage



\begin{figure}
  \begin{center}
   \includegraphics[width=0.8\textwidth]{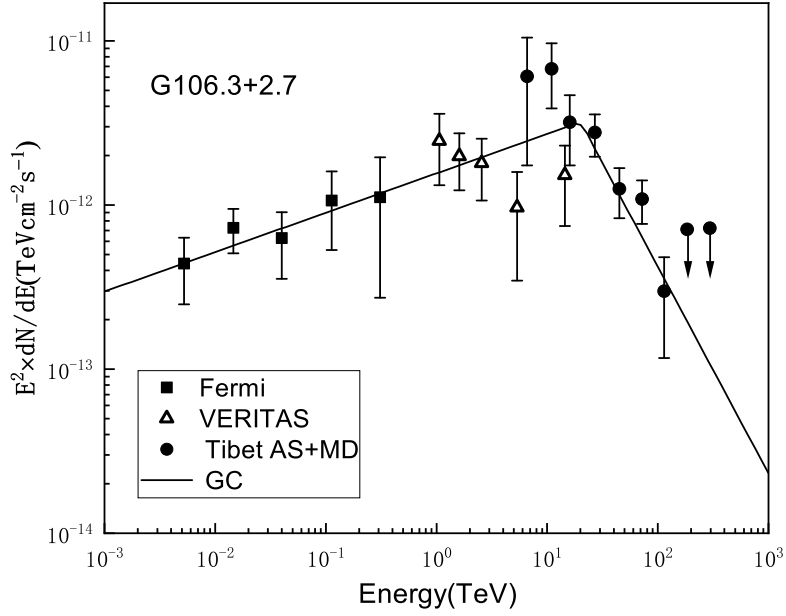} 
    \caption{ Predicted $\gamma$-ray spectra multiplied by $E^2_{\gamma}$
and comparisons with the SNR G106.3+2.7 spectrum [1,12,13]. The parameters are
$\Phi_0=1.2\times 10^{-17}~\rm TeV^{-1}cm^{-2}s^{-1}$, $\beta_{\gamma}=1.76$,
$\beta_p=1.25$ and $E_{\pi}^{GC}=20$ GeV, where the second break at $E_{\gamma}^{cut}$ is neglected
since we lack the data.
}\label{fig:1}
  \end{center}
\end{figure}



\end{thebibliography}
\end{document}